\documentclass[review,3p,11pt]{elsarticle}

\usepackage{lineno,hyperref}
\usepackage{xcolor}
\modulolinenumbers[5]

\usepackage{amsmath,amssymb}
\usepackage{graphicx}
\usepackage{epstopdf}
\usepackage{enumitem}

\usepackage{nomencl}
\usepackage{ifthen}
\renewcommand{\nomgroup}[1]{%
\ifthenelse{\equal{#1}{A}}{\item[\textit{Subscripts:}]}{%
\ifthenelse{\equal{#1}{B}}{\item[\textit{Subscripts:}]}{%
\ifthenelse{\equal{#1}{C}}{\item[\textit{Constants:}]}{}}}
}
\makenomenclature 

\journal{Journal of Process Control}

\makeindex 
\begin{document}

\begin{frontmatter}

\title{Electrolyte Flow Rate Control for Vanadium Redox Flow Batteries using the Linear Parameter Varying Framework\tnoteref{mytitlenote}}



\author{Ryan McCloy}
\ead{r.mccloy@unsw.edu.au}
\author{Yifeng Li\corref{steve}}
\ead{yifeng.li@voith.com}
\author{Jie Bao\corref{mycorrespondingauthor}}
\ead{j.bao@unsw.edu.au}
\author{Maria Skyllas-Kazacos}
\ead{M.Skyllas-Kazacos@unsw.edu.au}
\cortext[mycorrespondingauthor]{Corresponding author.}
\cortext[steven]{Y. Li is currently with the Voith Group, Germany.}
\address{School of Chemical Engineering, The University of New South Wales, Sydney, NSW 2052, Australia}

%
%

\begin{abstract}
In this article, an electrolyte flow rate control approach is developed for an all-vanadium redox flow battery (VRB) system based on the linear parameter varying (LPV) framework. The electrolyte flow rate is regulated to provide a trade-off between stack voltage efficiency and pumping energy losses, so as to achieve optimal battery energy efficiency. The nonlinear process model is embedded in a linear parameter varying state-space description and a set of state feedback controllers are designed to handle fluctuations in current during both charging and discharging. Simulation studies have been conducted under different operating conditions to demonstrate the performance of the proposed approach. This control approach was further implemented on a laboratory scale VRB system.
\end{abstract}


\end{frontmatter}

\linenumbers



\section{Introduction}
The all-vanadium redox flow battery (VRB) has attracted significant research interest, since it was invented by Skyllas-Kazacos and co-workers \cite{Skyllas-Kazacos1986, Skyllas-Kazacos1988} in the 1980s, largely due to its widely recognised potential for large scale energy storage applications. Despite its increasing interest, few contributions have focused on the development of targeted control system design as an approach to bringing to fruition the VRB's potential through improved performance and efficiency, particularly in dealing with time-varying charging/discharging power (or current).

Control of the electrolyte flow rate is important during VRB operation. While adequate flow rate is necessary to minimise concentration over potential losses and prevent side reactions, excess flow rate will lead to a higher pumping energy consumption. Implementation of variable flow rate has been proposed to reduce pumping losses and therefore increase the overall system efficiency, particularly when the battery is operated over a large range of state of charge (SOC) \cite{Skyllas-Kazacos1989,Blanc2009,tang2014}. Inspired by this idea, an experimental study on a laboratory scale system using a two stage pumping strategy was conducted \cite{Ma2012}, which suggested to maintain a low flow rate and only step up the flow rate towards the end of charging and discharging. Unlike in other battery areas (e.g., solid oxide fuel cells~\cite{Biao2013,VIJAY2019101}), very limited studies on advanced control and monitoring techniques for VRBs have been conducted. A simulation study on variable electrolyte flow rate control based on gain-scheduling was conducted and showed its potential benefit \cite{LiZhangBaoSkyllas17}. However, a real-time optimal electrolyte flow rate control approach has not been developed.

Dynamics of VRBs are highly nonlinear. Our previous studies have shown that a linear flow rate controller is unable to achieve satisfactory performance \cite{LiZhangBaoSkyllas17}. However computational complexity of nonlinear control algorithms is prohibitively high to VRB control and management systems. This motivates the proposed approach based on  a linear parameter varying (LPV) framework \cite{Sham88}. This is an attractive approach, as it offers a means of capturing the nonlinearity of the process for controller synthesis. In addition, this approach offers potentially minimal to no increase in online computational burden with respect to traditional fixed gain methods. A nonlinear model can be embedded in an LPV description by redefining the nonlinearities in the model as varying parameters. A common assumption is that these varying parameters are bounded, which allows for the system to be considered as varying within a polytope. Despite any conservatism introduced by LPV embedding, the LPV framework is advantageous due to simplification of the analysis and design of nonlinear control systems, by allowing for the application of powerful linear design tools to a wide range of nonlinear systems. From a system modelling perspective, an LPV embedding approach is also well suited to the problem of flow rate control in VRBs, since many of the performance metrics often used as reference targets, e.g., state of charge and electrolyte conversion factor, are dependent on the system state (i.e., ion concentrations), whose measurements are highly nonlinear. 

Consequently, this article details the modelling and optimal flow rate control of a VRB system to achieve efficient operation using the LPV framework. The paper is organised as follows: In Section~\ref{sec:sys_model}, the dynamic model of the VRB is developed and embedded in an LPV state space model representation. The synthesis of an LPV feedback tracking controller and closed-loop stability analysis are carried out in Section~\ref{sec:controller} via the use of a scheduling parameter. In Section~\ref{sec:implementation} the performance of the overall control scheme is studied through both simulations and experiments. Conclusions are drawn in Section~\ref{sec:conclusion}.

\printnomenclature

\section{Dynamic Model of the Vanadium Redox Battery}
\label{sec:sys_model}
\begin{figure}[tb]
	\centering
	\includegraphics[width=0.7\linewidth]{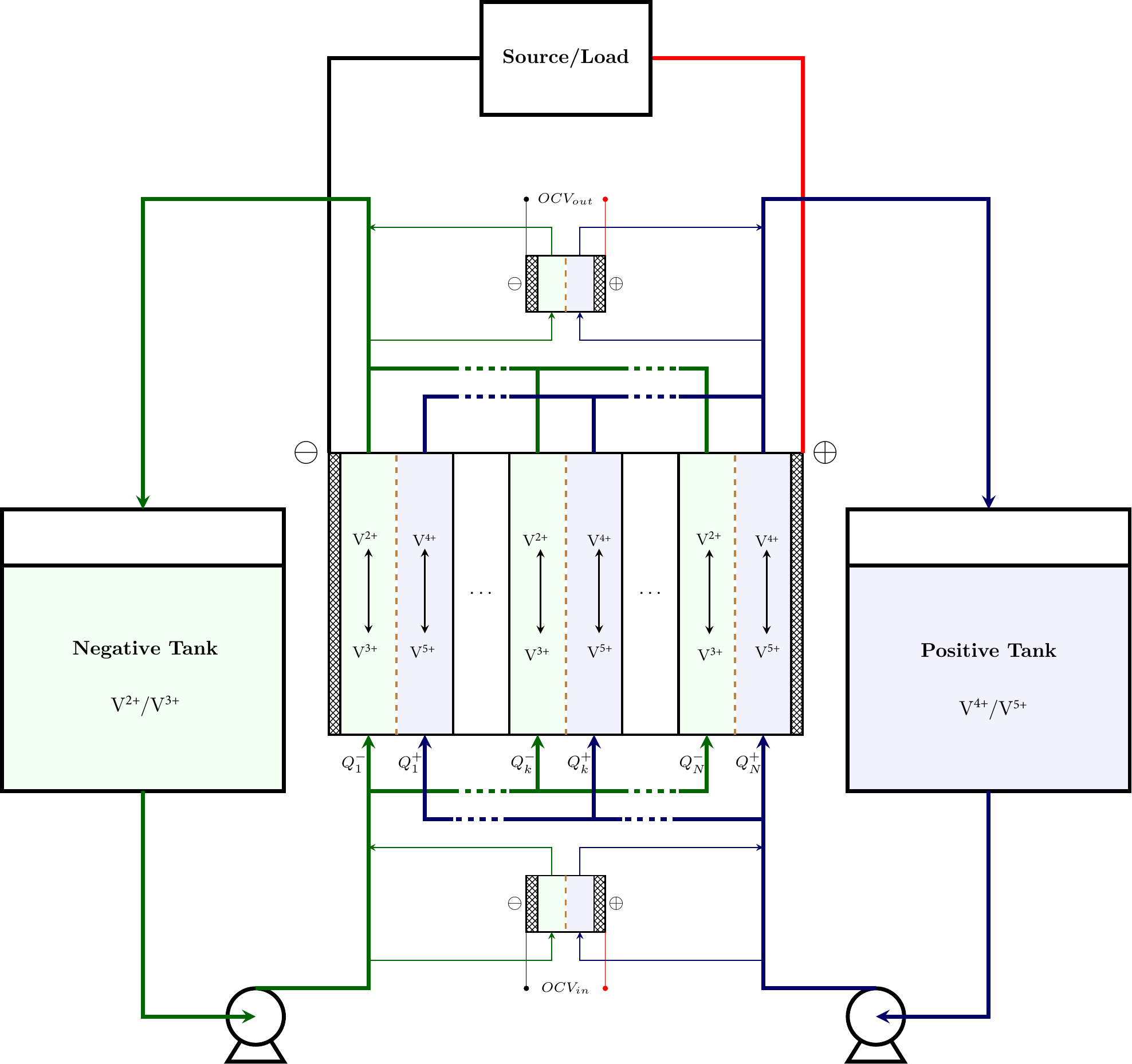}
	\caption{Schematic diagram of a multi-cell stack VRB system.}
	\label{fig:schematic}
\end{figure}

 As shown in Figure~\ref{fig:schematic}, a typical VRB system consists of: (1) a multiple cell battery stack, in which electrical-chemical energy conversion occurs; and (2) storage tanks (positive and negative) in which chemical energy is stored in the electrolytes. An ion exchange membrane separates the positive
and negative electrolytes in each cell. During operation, the electrolytes are continuously pumped through the cells (which are connected electrically in series and hydraulically in parallel). During charging, electrochemical reactions within the battery cells change the valence of the vanadium in the two electrolytes in the negative half-cell changing $V^{3+}$ to $V^{2+}$ and in the positive half-cell changing $V^{4+}$ ($VO^{2+}$) to $V^{5+}$ ($VO^+_2$) at certain conversion rate. This process is reversed during discharge. The main electrochemical reactions can be represented as follows:
	\begin{align}
		V^{3+}+e^-&\rightleftharpoons V^{2+} \label{reaction-}\\
		VO^{2+}+H_2O&\rightleftharpoons VO_2^++2H^++e^- \label{reaction+}
	\end{align}
The battery SOCs, before and after the cell stacks, can be estimated from measurements of the voltages of open circuit cells (OCVs) at both the inlet and outlet of the cells. 

\subsection{Dynamic Mass Balance Model of the VRB}
The dynamics of the vanadium ion concentrations of $V^{2+}$, $V^{3+}$, $V^{4+}$ ($VO^{2+}$) and $V^{5+}$ ($VO^+_2$) in the battery cell, denoted as $c_{c,2}$, $c_{c,3}$, $c_{c,4}$ and $c_{c,5}$ respectively, together with the vanadium concentrations in the tank, denoted as $c_{t,2}$, $c_{t,3}$, $c_{t,4}$ and $c_{t,5}$, can be modelled as follows \cite{tang2014}:
\begin{equation}
\begin{split}
\label{eq:concentrations}
	\frac{d }{dt}c_{c,2} &= -\frac{1}{W_{pe}}\left(\frac{k_2}{d}c_{c,2} + \frac{k_4}{d}c_{c,4} + 2\frac{k_5}{d}c_{c,5}\right) 
					 + \frac{1}{ML_{pe}W_{pe}H_{pe}}(c_{t,2}-c_{c,2})Q + \frac{1}{nL_{pe}W_{pe}H_{pe}F}I,\\
	\frac{d}{dt}c_{c,3} &= -\frac{1}{W_{pe}}\left(\frac{k_3}{d}c_{c,3} - 2\frac{k_4}{d}c_{c,4} - 3\frac{k_5}{d}c_{c,5}\right)  + \frac{1}{ML_{pe}W_{pe}H_{pe}}(c_{t,3}-c_{c,3})Q - \frac{1}{nL_{pe}W_{pe}H_{pe}F}I,\\
	\frac{d }{dt}c_{c,4} &= -\frac{1}{W_{pe}}\left(- 3\frac{k_2}{d}c_{c,2} - 2\frac{k_3}{d}c_{c,3} + \frac{k_4}{d}c_{c,4}\right) + \frac{1}{ML_{pe}W_{pe}H_{pe}}(c_{t,4}-c_{c,4})Q - \frac{1}{nL_{pe}W_{pe}H_{pe}F}I,\\
	\frac{d }{dt}c_{c,5} &= -\frac{1}{W_{pe}}\left(2\frac{k_2}{d}c_{c,2} + \frac{k_3}{d}c_{c,3} + \frac{k_5}{d}c_{c,5}\right)  + \frac{1}{ML_{pe}W_{pe}H_{pe}}(c_{t,5}-c_{c,5})Q + \frac{1}{nL_{pe}W_{pe}H_{pe}F}I,\\
		\frac{d }{dt}c_{t,2} &= \frac{1}{V_t}(c_{c,2}-c_{t,2})Q,\\
	\frac{d }{dt}c_{t,3} &= \frac{1}{V_t}(c_{c,3}-c_{t,3})Q,\\
	\frac{d }{dt}c_{t,4} &= \frac{1}{V_t}(c_{c,4}-c_{t,4})Q,\\
	\frac{d }{dt}c_{t,5} &= \frac{1}{V_t}(c_{c,5}-c_{t,5})Q,\\
\end{split}
\end{equation}
where $Q=Q(t)$ is the pump flow rate (controlled input) and $I=I(t)$ is the electrical current (treated as a measured disturbance input). Parameters $k_2$, $k_3$, $k_4$ and $k_5$ are the diffusivity coefficients for $V^{2+}$, $V^{3+}$, $V^{4+}$ and $V^{5+}$ across the membrane respectively. $L_{pe}$, $W_{pe}$ and $H_{pe}$ are the length, width and height of the porous electrode respectively and $d$ is the thickness of membrane. $V_t$ denotes the volume of electrolyte in each half cell, $F$ is Faraday's constant, $n$ is the number of electrons transferred in the reaction, and $M$ is the number of cells in the stack.

We consider two sensors, providing measurements of the inlet and outlet open circuit cell voltage, $E^{IN}_{OCV}$ and $E^{OUT}_{OCV}$ respectively. Based on the Nernst Equation, these measurements can be represented as functions of the vanadium concentrations as shown below:
\begin{equation}
\label{eq:OCVs}
E^{IN}_{OCV} =  E'_0 + \frac{R T}{n F} \ln \left(\frac{c_{t,2}c_{t,5}}{c_{t,3}c_{t,4}}\right), \qquad E^{OUT}_{OCV} = E'_0 + \frac{R T}{n F} \ln\left(\frac{c_{c,2}c_{c,5}}{c_{c,3}c_{c,4}}\right),
\end{equation}
where $E'_0$ is the formal potential, $R$ is the Gas constant (J mol$^{-1}$ K$^{-1}$) and $T$ is the temperature.
The concentrations, $c$, pump flow rate, $Q$, and electrical current, $I$, are considered bounded as
\begin{equation}
\label{eq:sysbounds}
c_{min} \leq c \leq c_{max}, \qquad Q_{min} \leq Q \leq Q_{max}, \qquad I_{min} \leq I \leq I_{max}.
\end{equation}

\subsection{LPV Embedded State Space Model}\label{sec:lpv_model}
Intuitively, and without loss of generality, the dynamics of a VRB can be understood as a linear time-varying model, which is dependent on the variation of the concentrations (and hence SOC) of the system. In a practical sense, this observation motivates the following linear parameter varying approach, in which the varying parameters are naturally functions of the concentrations.
To embed the nonlinear VRB model of~\eqref{eq:concentrations} into an LPV state space description, we begin by defining the states, $x(t)= (x_1, x_2)$ as
\begin{equation}
\label{eq:states}
	x_1 := \frac{c_{t,2}c_{t,5}}{c_{t,3}c_{t,4}},\qquad	x_2 := \frac{c_{c,2}c_{c,5}}{c_{c,3}c_{c,4}}.
\end{equation}
This particular definition for the states is attractive, since from~\eqref{eq:OCVs} and recalling the positive, non-zero bound on the concentrations (see~\eqref{eq:sysbounds}),  the states, $x_1$, $x_2$, in~\eqref{eq:states} can be reconstructed from the OCVs and formal potential explicitly:
\begin{equation}
\label{eq:OCVtox}
x_1 = e^{\frac{nF}{RT}(E^{IN}_{OCV} - E'_0)}, \qquad x_2 =  e^{\frac{nF}{RT}(E^{OUT}_{OCV} - E'_0)}.
\end{equation}
Taking the derivative of each state in~\eqref{eq:states} yields
\begin{equation}
\begin{split}
\label{eq:state_derivative}
	\frac{d}{dt}x_1 &= \frac{c_{t,5}}{c_{t,3}c_{t,4}}\frac{d}{dt}c_{t,2} 
				- \frac{c_{t,2}c_{t,5}}{c_{t,3}^2c_{t,4}}\frac{d}{dt}c_{t,3} 
				- \frac{c_{t,2}c_{t,5}}{c_{t,3}c_{t,4}^2}\frac{d}{dt}c_{t,4} 
				+ \frac{c_{t,2}}{c_{c,3}c_{c,4}}\frac{d}{dt}c_{t,5},\\
	\frac{d}{dt}x_2 &= \frac{c_{c,5}}{c_{c,3}c_{c,4}}\frac{d}{dt}c_{c,2} 
				- \frac{c_{c,2}c_{c,5}}{c_{c,3}^2c_{c,4}}\frac{d}{dt}c_{c,3} 
				- \frac{c_{c,2}c_{c,5}}{c_{c,3}c_{c,4}^2}\frac{d}{dt}c_{c,4} 
				+ \frac{c_{c,2}}{c_{c,3}c_{c,4}}\frac{d}{dt}c_{c,5}.
\end{split}
\end{equation}
Substituting~\eqref{eq:concentrations} into~\eqref{eq:state_derivative} gives
\begin{equation}
\begin{split}
\label{eq:state_derivative_sub}
	\frac{d}{dt}x_1 &= \frac{1}{V_t}\left(\frac{c_{t,5}}{c_{t,3}c_{t,4}}(c_{c,2}-c_{t,2}) 
				- \frac{c_{t,2}c_{t,5}}{c_{t,3}^2c_{t,4}}(c_{c,3}-c_{t,3}) \right. \dots \\
				&\qquad \left. - \frac{c_{t,2}c_{t,5}}{c_{t,3}c_{t,4}^2}(c_{c,4}-c_{t,4})
				+ \frac{c_{t,2}}{c_{c,3}c_{c,4}}(c_{c,5}-c_{t,5})\right)Q,\\
	\frac{d}{dt}x_2 &= -\frac{1}{W_{pe}}\left(\frac{c_{c,5}}{c_{c,3}c_{c,4}}\left(\frac{k_2}{d}c_{c,2} + \frac{k_4}{d}c_{c,4} + 2\frac{k_5}{d}c_{c,5}\right) \right. \dots \\
							     &\qquad \left. - \frac{c_{c,2}c_{c,5}}{c_{c,3}^2c_{c,4}}\left(\frac{k_3}{d}c_{c,3} - 2\frac{k_4}{d}c_{c,4} - 3\frac{k_5}{d}c_{c,5}\right) \right.\dots \\
							    &\qquad\left. - \frac{c_{c,2}c_{c,5}}{c_{c,3}c_{c,4}^2}\left(- 3\frac{k_2}{d}c_{c,2} - 2\frac{k_3}{d}c_{c,3} + \frac{k_4}{d}c_{c,4}\right) \right. \dots \\
							    &\qquad \left. + \frac{c_{c,2}}{c_{c,3}c_{c,4}}\left(2\frac{k_2}{d}c_{c,2} + \frac{k_3}{d}c_{c,3} + \frac{k_5}{d}c_{c,5}\right)\right) \dots \\
				&\quad+\frac{1}{ML_{pe}W_{pe}H_{pe}} \left(\frac{c_{c,5}}{c_{c,3}c_{c,4}}(c_{t,2}-c_{c,2})
										- \frac{c_{c,2}c_{c,5}}{c_{c,3}^2c_{c,4}}(c_{t,3}-c_{c,3}) \right. \dots \\
										&\qquad \left. - \frac{c_{c,2}c_{c,5}}{c_{c,3}c_{c,4}^2}(c_{t,4}-c_{c,4})
										+ \frac{c_{c,2}}{c_{c,3}c_{c,4}}(c_{t,5}-c_{c,5}) \right)Q \dots \\
				&\quad+ \frac{1}{nL_{pe}W_{pe}H_{pe}F} \left(\frac{c_{c,5}}{c_{c,3}c_{c,4}}
											+ \frac{c_{c,2}c_{c,5}}{c_{c,3}^2c_{c,4}}
											+ \frac{c_{c,2}c_{c,5}}{c_{c,3}c_{c,4}^2}
											+ \frac{c_{c,2}}{c_{c,3}c_{c,4}} \right)I			
\end{split}
\end{equation}

By defining the varying parameters, $\rho_1,\rho_2,\rho_3,\rho_4$, as
\begin{equation}
\begin{split}
\label{eq:rho}
\rho_1 &= \frac{1}{V_t}\left(\frac{c_{t,5}}{c_{t,3}c_{t,4}}(c_{c,2}-c_{t,2}) 
				- \frac{c_{t,2}c_{t,5}}{c_{t,3}^2c_{t,4}}(c_{c,3}-c_{t,3}) \right. \dots \\
				&\qquad \left. - \frac{c_{t,2}c_{t,5}}{c_{t,3}c_{t,4}^2}(c_{c,4}-c_{t,4})
				+ \frac{c_{t,2}}{c_{c,3}c_{c,4}}(c_{c,5}-c_{t,5})\right),\\
\rho_2 &= -\frac{1}{W_{pe}}\left( \left(\frac{k_2}{d} + \frac{k_4}{d}\frac{c_{c,4}}{c_{c,2}} + 2\frac{k_5}{d}\frac{c_{c,5}}{c_{c,2}}\right) 
					-\left(\frac{k_3}{d} - 2\frac{k_4}{d}\frac{c_{c,4}}{c_{c,3}} - 3\frac{k_5}{d}\frac{c_{c,5}}{c_{c,3}}\right) \right.\dots\\
					&\qquad\left. -\left(- 3\frac{k_2}{d}\frac{c_{c,2}}{c_{c,4}} - 2\frac{k_3}{d}\frac{c_{c,3}}{c_{c,4}} + \frac{k_4}{d}\right)
					+\left(2\frac{k_2}{d}\frac{c_{c,2}}{c_{c,5}} + \frac{k_3}{d}\frac{c_{c,3}}{c_{c,5}} + \frac{k_5}{d}\right) \right),\\
\rho_3 &= \frac{1}{ML_{pe}W_{pe}H_{pe}} \left(\frac{c_{c,5}}{c_{c,3}c_{c,4}}(c_{t,2}-c_{c,2})
							- \frac{c_{c,2}c_{c,5}}{c_{c,3}^2c_{c,4}}(c_{t,3}-c_{c,3}) \right. \dots \\
							&\qquad \left. - \frac{c_{c,2}c_{c,5}}{c_{c,3}c_{c,4}^2}(c_{t,4}-c_{c,4})
							+ \frac{c_{c,2}}{c_{c,3}c_{c,4}}(c_{t,5}-c_{c,5}) \right),\\
\rho_4 &=  \frac{1}{nL_{pe}W_{pe}H_{pe}F} \left(\frac{c_{c,5}}{c_{c,3}c_{c,4}}
							+ \frac{c_{c,2}c_{c,5}}{c_{c,3}^2c_{c,4}}
							+ \frac{c_{c,2}c_{c,5}}{c_{c,3}c_{c,4}^2}
							+ \frac{c_{c,2}}{c_{c,3}c_{c,4}} \right),
\end{split}
\end{equation}
we can then parameterise the system~\eqref{eq:state_derivative_sub} in an LPV state space representation\footnote{Note that this parameterisation is not unique and has been chosen to address specific controller design prerequisites (see Section~\ref{sec:controller}).}
\begin{equation}
\label{eq:cont_ss}
\frac{d}{dt}x(t) = \begin{bmatrix} 0&0\\0&\rho_2 \end{bmatrix}x(t) + \begin{bmatrix} \rho_1\\ \rho_3 \end{bmatrix}u(t) + \begin{bmatrix} 0\\ \rho_4 \end{bmatrix}w(t),
\end{equation}
where $u(t) := Q(t)$ and $w(t) := I(t)$.

If we consider an Euler discretisation with sampling period $\tau$ for a particular system described by some continuous function $\dot{x} = f(x,u)$ to be equal to $x^+ = x + \tau f(x,u)$, then the following provides a discrete-time LPV model for the system in~\eqref{eq:cont_ss} 
\begin{equation}
\label{eq:disc_ss}
x(k+1) = \underbrace{\begin{bmatrix} 1&0\\0&1+\tau \rho_2 \end{bmatrix}}_{A(\rho)}x(k) + \underbrace{\begin{bmatrix} \tau \rho_1\\ \tau \rho_3 \end{bmatrix}}_{B(\rho)}u(k) + \underbrace{\begin{bmatrix} 0\\ \tau \rho_4 \end{bmatrix}}_{E(\rho)}w(k),
\end{equation}

The dynamic system model in~\eqref{eq:concentrations} has been embedded in an LPV state space description, whereby the varying parameters $\rho_1,\dots,\rho_4$ are dependent on the states of the system. In our practical setup we can assume that the varying parameter is measured as described in the following. 
\subsection{Varying Parameter Update from Measurements}
\label{app:ideal_parameter}
We assume that the two half cells of the battery are fully balanced and provide details for reconstructing the varying parameters from measurements based on the following model: (in practice, the SOC imbalances can be caused by the crossover of vanadium ions which can be monitored and corrected by remixing the positive and negative electrolyte) 
\begin{equation}
\label{eq:ideal}
c_2 + c_3 = c_4 + c_5 =\bar{c} \qquad \text{and} \qquad \frac{c_2}{c_2+c_3} = \frac{c_5}{c_5+c_4},
\end{equation}
where $\bar{c}$ is the total concentration. From~\eqref{eq:states} and~\eqref{eq:ideal} we have:
\begin{equation}
\label{eq:balanced}
\frac{c_{t,2}^2}{(\bar{c}-c_{t,2})^2} = x_1, \qquad \frac{c_{c,2}^2}{(\bar{c}-c_{c,2})^2} = x_2, \qquad c_5=c_2, c_3=c_4,
\end{equation}
which can be solved explicitly to give
\begin{equation}
\begin{split}
\label{eq:cs}
c_{t,5} = c_{t,2} = \frac{\bar{c}(\sqrt{x_1} - x_1)}{1-x_1}, \qquad c_{t,3} = c_{t,4} = \bar{c}- c_{t,2},\\
c_{c,5} = c_{c,2} = \frac{\bar{c}(\sqrt{x_2} - x_2)}{1-x_2}, \qquad c_{c,3} = c_{c,4} = \bar{c}- c_{c,2},\\
\end{split}
\end{equation}
Hence, using~\eqref{eq:cs}, the available OCV measurements~\eqref{eq:OCVs} and the definitions for the varying parameters~\eqref{eq:rho}, \eqref{eq:rho5} we can update ``measurements'' of $\rho$ at each time instant. In the case when the varying parameter is not exactly known but measured with bounded errors, techniques similar to those used in~\cite{NazariSeronDeDona15} could be employed.

\section{Control Design}
\label{sec:controller}
The objective is to manipulate the electrolyte flow rate, $u$, to achieve a desired conversion rate (fraction conversion per pass), $X_s$, subject to a charging current, $w$. 
The (fraction) conversion per pass refers to the fractional change of the SOC of the electrolyte every time it passes through the cell stack and is a function of charging/discharging current, electrolyte flow rate and the battery SOC. Achieving a suitable conversion per pass will lead to  efficient battery operation \cite{tang2014, LiZhangBaoSkyllas17}.

In this section we will augment the system model of Section~\ref{sec:sys_model} to permit offset-free tracking and design a scheduled state-feedback controller to achieve this purpose. In addition, due to the hardware limitations of a pilot VRB battery experimental control study, a low computational burden approach is desired, and hence a convex polytopic control method, leveraging the LPV embedding of Section~\ref{sec:sys_model}, is developed in the following. 
\subsection{Augmented Model}
Here we provide details for explicitly defining the error in conversion factor as a function of the states via a model for the state of charge. The inlet OCV can be described by (see~\cite{LiZhangBaoSkyllas17})
\begin{equation}
\label{eq:OCVSOC}
E^{IN}_{OCV} =  E'_0 + 2\frac{R T}{n F} \ln \left(\frac{SOC}{1-SOC}\right).
\end{equation}
From~\eqref{eq:OCVs},\eqref{eq:states} and~\eqref{eq:OCVSOC} and noting the positive, non-zero bound on $x_1$ (see~\eqref{eq:sysbounds},\eqref{eq:states}) we then have
\begin{equation}
\label{eq:SOCx}
SOC = \frac{\sqrt{x_1}}{1+\sqrt{x_1}}.
\end{equation}
The relationship between the inlet and outlet open circuit cell voltages, $E^{IN}_{OCV}$ and $E^{OUT}_{OCV}$~\eqref{eq:OCVs}, and the state of charge, $SOC$, can be expressed as 
\begin{equation}
\begin{split}
\label{eq:deltaOCV}
E^{OUT}_{OCV} - E^{IN}_{OCV} &= \frac{R T}{n F} \ln \left(\frac{c_{c,2}c_{c,5}}{c_{c,3}c_{c,4}}\frac{c_{t,3}c_{t,4}}{c_{t,2}c_{t,5}}\right)\\
&=2\frac{R T}{n F} \ln  \left(\frac{SOC + X_c - X_c\cdot SOC}{SOC - X_c\cdot SOC}\right) \\
&=2\frac{R T}{n F} \ln  \left(\frac{1-SOC -X_d+X_d\cdot SOC}{1-SOC+X_d\cdot SOC}\right),
\end{split}
\end{equation}
where $X_c$ and $X_d$ denote the conversion per pass during charging and discharging respectively. Substitution of $x_1$, $x_2$~\eqref{eq:states} in~\eqref{eq:deltaOCV} and noting their positive, non-zero bounds yields
\begin{equation}
\begin{split}
\label{eq:SOCx1x2}
\sqrt{\frac{x_2}{x_1}} &=\frac{SOC + X_c - X_c\cdot SOC}{SOC - X_c\cdot SOC} \\
&=\frac{1-SOC -X_d+X_d\cdot SOC}{1-SOC+X_d\cdot SOC}.
\end{split}
\end{equation}
Denote the conversion per pass as $X$. Following substitution of~\eqref{eq:SOCx} in~\eqref{eq:SOCx1x2} and suitable rearrangement, we obtain
\begin{equation}
\label{eq:Xfactor}
X=\begin{cases}
X_c = 1 - \frac{1+\sqrt{x_1}}{1+\sqrt{x_2}}, & \text{during charging} \\
X_d = \frac{1 - \sqrt{\frac{x_2}{x_1}}}{1+\sqrt{x_2}},  & \text{during discharging}
\end{cases}
\end{equation}
 The objective can then be formulated as a tracking control problem, whereby the flow rate or input, $u$, is manipulated such that the tracking error, $e(k) = X_s-X(k)$ is driven to zero
for a desired conversion factor $X_s$.

 We then define a ``performance'' output, $y_p = X_c$  or $y_p = X_d$ during charging or discharging respectively, as
\begin{equation}
\label{eq:y}
y_p = \underbrace{\begin{bmatrix}  \rho_5 & 0\end{bmatrix}}_{C(\rho)}x,
\end{equation}
where, 
\begin{equation}
\label{eq:rho5}
\rho_{5,c} = \frac{\sqrt{x_2}-\sqrt{x_1}}{x_1(1+\sqrt{x_2})}, \qquad \rho_{5,d} =  \frac{1 - \sqrt{\frac{x_2}{x_1}}}{x_1(1+\sqrt{x_2})},
\end{equation}
with $\rho_{5,c}$, $\rho_{5,d}$ corresponding to a respective conversion rate, $X_c$, $X_d$ in~\eqref{eq:Xfactor}.

In order to achieve asymptotic tracking of a reference, $r = X_s$, we can add an integral action state, $\sigma$, to~\eqref{eq:disc_ss}, satisfying
\begin{equation}
\label{eq:z}
\sigma(k+1) = \sigma +\tau(r - y_p).
\end{equation}
Hence, we define an augmented state, $\zeta := \begin{bmatrix} x \\ \sigma \end{bmatrix}$ satisfying
\begin{equation}
\label{eq:zeta}
\zeta(k+1) =\underbrace{\begin{bmatrix} A(\rho) & 0 \\ - \tau C(\rho) & 1 \end{bmatrix}}_{A_\zeta} \zeta 
+ \underbrace{\begin{bmatrix} B(\rho) \\ 0 \end{bmatrix}}_{B_\zeta}u 
+ \begin{bmatrix} E(\rho) \\ 0 \end{bmatrix}w
+ \begin{bmatrix} 0 \\ \tau \end{bmatrix}r,
\end{equation}
where the $0$ matrices are of appropriate dimension. Using the augmented LPV state space model of~\eqref{eq:zeta} we can now design a state feedback controller to achieve offset-free tracking.
\subsection{Scheduled State Feedback Design}
\label{sec:param}
\begin{figure}
\centering
  \includegraphics[width=0.6\linewidth]{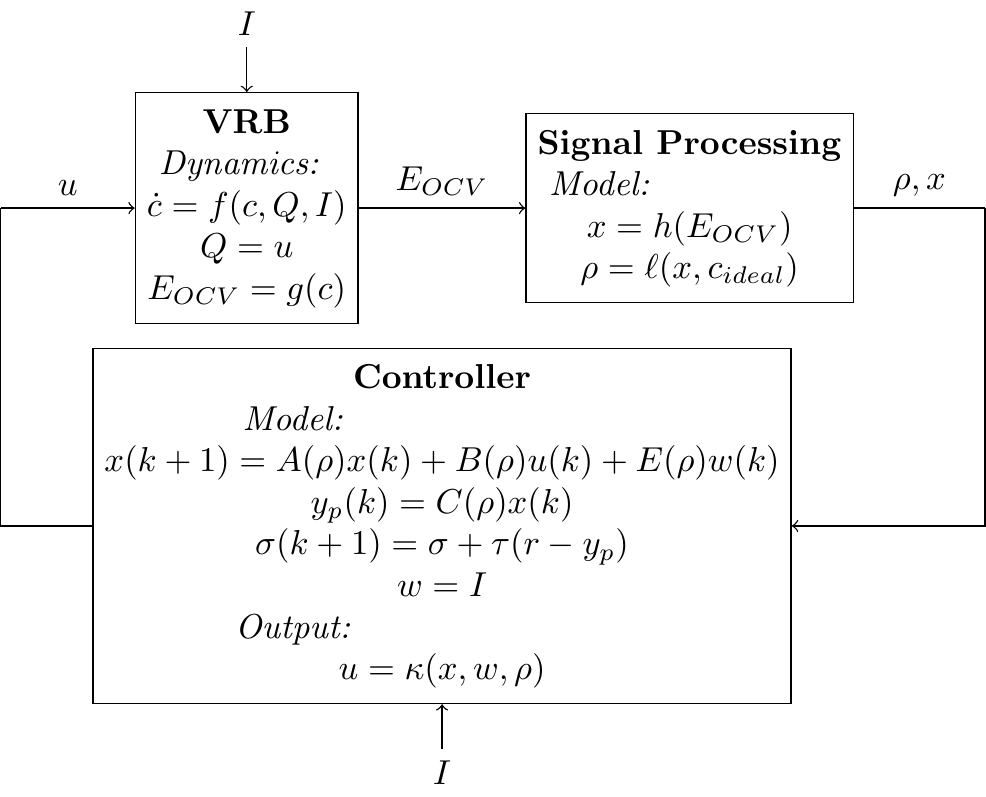}
  \caption{VRB Control System Block Diagram}
  \label{fig:block_diagram}
\end{figure}

A simplified block diagram for the overall control scheme is shown in Figure~\ref{fig:block_diagram}. 
We employ the augmented state-feedback tracking controller with disturbance accommodation:
\begin{equation}
\label{eq:u}
u(k) = u^*(k) -K_x(k)(x(k)-x^*(k)) -K_\sigma(k)\sigma -K_w(k)w(k),
\end{equation}
where $x^*$ is the desired state value and $u^*$ is the corresponding control input. The feedback gain, $K_w(k) = K_w(\rho(k)) = B^+\!(\rho(k)) E(\rho(k))$ in~\eqref{eq:u}, is a standard implementation of disturbance accommodation, where $B^+$ denotes the Moore-Penrose pseudo-inverse of $B$. The feedback gains, $K_x(k) = K_x(\rho(k)), K_\sigma(k) = K_\sigma(\rho(k))$, are designed to achieve certain performance characteristics and are defined as 
\begin{equation}
\label{eq:KxK}
\begin{bmatrix} K_x & K_\sigma \end{bmatrix} := K_\zeta,
\end{equation}
which is an LQR for the augmented system~\eqref{eq:zeta}, provided $(A_\zeta(\rho),B_\zeta(\rho))$ is stabilisable (i.e., \textit{controllable}, see Section~\ref{sec:analysis}) for all $\rho(k)$. By updating measurements of $\rho = \rho(k)$, and hence $(A_\zeta(\rho),B_\zeta(\rho))$, at each time step $k$, the feedback gain, $K_\zeta = K_\zeta(\rho(k))$ in \eqref{eq:KxK}, can be computed as
\begin{equation}\label{eq:Krhok}
    K_\zeta = (R + B_\zeta^\top \! P B_\zeta)^{-1} B_\zeta^\top \! P A_\zeta,
\end{equation}
where $P = P(\rho(k))$ is the positive definite solution to the discrete-time algebraic Riccati equation
\begin{equation}\label{eq:Riccati}
    A_\zeta^\top \! P A_\zeta - (A_\zeta^\top \! P B_\zeta)(R + B_\zeta^\top \! P B_\zeta)^{-1} (B_\zeta^\top \! P A_\zeta) + Q - P = 0
\end{equation}
and hence, $u_{\zeta}(k) = u_{\zeta}^*(k) - K_{\zeta} (\zeta(k)-\zeta^*(k))$ (from~\eqref{eq:u}) minimises, at each time step $k$, the quadratic cost function (i.e., performance index)
\begin{equation}
\label{eq:cost}
J(u_{\zeta}) = \sum^\infty_{k=1} \left (   (\zeta-\zeta^*)^\top Q  (\zeta-\zeta^*) + (u_{\zeta}-u_{\zeta}^*)^\top R (u_{\zeta}-u_{\zeta}^*)\right),
\end{equation}
where $Q = Q(\rho(k))$ and $R = R(\rho(k))$ are parameter dependent state and control penalty weightings respectively. It is important to note here that the state weighting, $Q$, which directly affects the tracking performance $X \to X_s$, is in fact a reflection of the physical VRB stack efficiency. In addition, suppose the efficiency (in terms of power consumption) of the pump is known, then this information can be incorporated into the selection of the control weighting, $R$. Hence, the cost function in~\eqref{eq:cost} can be understood as a trade-off between stack efficiency, battery capacity and pump energy consumption, through relative changes in $Q$ and $R$ respectively.

In light of hardware limitations, one might consider the relatively straightforward task of computing the LQR feedback gain, $K_\zeta(\rho(k))$ in \eqref{eq:Krhok}, via solution to the algebraic Riccati equation \eqref{eq:Riccati}, at each time step, may be computationally infeasible. As a significantly less exhaustive alternative, we can embed the LPV system \eqref{eq:disc_ss} into a convex polytopic description and form the feedback gains as linear combinations of those corresponding to the vertices of the system polytopes. This requires the very reasonable assumption that the varying parameter, defined as 
\begin{equation}\label{eq:rhov}
    \rho := (\rho_1,\rho_2,\rho_3,\rho_4,\rho_5),
\end{equation}
is bounded  as
\begin{equation}
\label{eq:rho_constraint}
\rho_{min} \leq \rho \leq \rho_{max},
\end{equation}
which can be found experimentally by considering each element of $\rho$ (e.g., $\rho_{1,min} \leq \rho_1 \leq \rho_{1,max}$), for the purpose of controller design. Under normal operation this will force much tighter bounds than those imposed by physical system constraints. Then, the parameter $\rho$ lies in a bounded set (i.e., convex polytope) $\mathcal{P} \in \mathbb{R}^L$, with $L=5$ (see \eqref{eq:rhov}), and thus the system matrices in~\eqref{eq:disc_ss} and \eqref{eq:y} can be written in the following polytopic form:
\begin{equation}
\label{eq:sscomb}
A(\rho) = \sum^{N}_{j=1} \xi_j(\rho)A_j, \qquad B(\rho) = \sum^{N}_{j=1} \xi_j(\rho)B_j, \qquad C(\rho) = \sum^{N}_{j=1} \xi_j(\rho)C_j, \qquad E(\rho) = \sum^{N}_{j=1} \xi_j(\rho)E_j,
\end{equation}
given constant matrices $A_j$, $B_j$, $C_j$, $E_j$ of compatible dimension. The number of vertices is denoted by $N=2^L=32$ and the functions $\xi_j : \mathcal{P} \to \mathbb{R}$ satisfy 
$\xi_j(\rho) \geq 0$, and $\sum^{N}_{j=1} \xi_j(\rho) = 1$, 
for all $\rho \in \mathcal{P}$. Hence, for each $\rho \in \mathcal{P}$, $(A(\rho),B(\rho),C(\rho),E(\rho))$ lies in the convex hull $ \mathrm{Co}\{$ $(A_1,B_1,C_1,E_1)$, $(A_2,B_2,C_2,E_2)$, $\dots$,
$(A_{N},B_{N},C_{N},E_{N})$ $\}$. 
In order to explicitly handle the dimension of $\rho$ \eqref{eq:rhov}, such that the mapping $\mathcal{P} \to \mathbb{R}$ is possible, and hence the combinations of matrices, e.g., $\xi_j A_j$ \eqref{eq:sscomb}, are possible, we need to express the functions $\xi_j(\rho)$ as a normalised linear combination of the varying parameter $\rho$. Hence, if we consider all combinations of the maximum and minimum values for the elements of $\rho(k)$ (see~\eqref{eq:rho_constraint}), then we can define $\xi_j(\rho) := \xi_j(\rho(k)) := \xi_j(k)$ as
\begin{equation}
\begin{split}
\label{eq:xis}
\xi_1(k) &= \phi_1(k) \phi_2(k) \cdot \ldots \cdot \phi_L(k) ,\\
\xi_2(k) &= (1-\phi_1(k)) \phi_2(k) \cdot \ldots \cdot \phi_L(k) ,\\
&\,\,\,\,\vdots\\
\xi_{N-1}(k) &= (1-\phi_1(k)) (1-\phi_2(k)) \cdot \ldots \cdot (1-\phi_{L-1}(k)) \phi_L(k),\\
\xi_{N}(k) &= (1-\phi_1(k)) (1-\phi_2(k)) \cdot \ldots \cdot (1-\phi_{L}(k)),
\end{split}
\end{equation}
where
\begin{equation}
\label{eq:xixy}
\phi_i(k) \triangleq \frac{\rho_{i,max} - \rho_i(k)}{\rho_{i,max}-\rho_{i,min}}, \qquad \text{for} \quad i = 1,\dots, L.
\end{equation}
Then, using~\eqref{eq:disc_ss},~\eqref{eq:rho_constraint}--\eqref{eq:xixy}, the state matrices can be computed as a convex combination of their vertices, where
\begin{equation}
\begin{split}
\label{eq:Arhos}
A_1 &= A(\rho_{1,min},\rho_{2,min},\dots,\rho_{L,min}), \\
A_2 &= A(\rho_{1,max},\rho_{2,min},\dots,\rho_{L,min}), \\
&\vdots \\
A_{N-1} &=A(\rho_{1,max},\rho_{2,max},\dots,\rho_{L-1,max},\rho_{L,min}) \\
A_N &=A(\rho_{1,max},\rho_{2,max},\dots,\rho_{L,max})
\end{split}
\end{equation}
A similar treatment can be performed for the matrices $B(\rho)$, $C(\rho)$ and $E(\rho)$ in~\eqref{eq:disc_ss} and hence $A_\zeta(\rho)$, $B_\zeta(\rho)$ in \eqref{eq:zeta}.

Provided the pairs $(A_{\zeta,j},B_{\zeta,j})$ (formed analogously to \eqref{eq:sscomb} given \eqref{eq:zeta}) are stabilisable, for $j=1,\dots,N$, feedback gains are then implemented as (see, e.g.,~\cite{McCloyDeDonaSeron18})  
\begin{equation}
\label{eq:Kcomb}
K_\zeta =  \sum^{N}_{j=1} \xi_j(\rho)K_{\zeta,j}, \qquad K_w = \sum^{N}_{j=1} \xi_j(\rho)K_{w,j},
\end{equation}
where the vertices are computed as (cf.~\eqref{eq:Krhok})
\begin{equation}
\label{eq:Kvert}
K_{\zeta,j} = (R_j + B_{\zeta,j}^\top P_j B_{\zeta,j})^{-1} B_{\zeta,j}^\top P_j A_{\zeta,j}, \qquad K_{w,j} = B_j^+ E_j,
\end{equation}
where $B_j^+$ denotes the Moore-Penrose pseudo-inverse of $B_j$, and $P_j$ is the solution to \eqref{eq:Riccati} at the system vertices $(A_{\zeta,j},B_{\zeta,j})$ given the weights $Q_j$ and $R_j$, such that
\begin{equation}
\begin{split}
\label{eq:Krhos}
K_{\zeta,1} &= K_\zeta(\rho_{1,min},\rho_{2,min},\dots,\rho_{L,min}), \\
K_{\zeta,2} &= K_\zeta(\rho_{1,max},\rho_{2,min},\dots,\rho_{L,min}), \\
&\vdots \\
K_{\zeta,N-1} &= K_\zeta(\rho_{1,max},\rho_{2,max},\dots,\rho_{L-1,max},\rho_{L,min}) \\
K_{\zeta,N} &= K_\zeta(\rho_{1,max},\rho_{2,max},\dots,\rho_{L,max})
\end{split}
\end{equation}
The final control law, which implements the feedback gains via convex combination, is then given from~\eqref{eq:u}--\eqref{eq:Krhos} as
\begin{equation}
\label{eq:ucomb}
u(k) := u_{\zeta}(k) = u_{\zeta}^*(k) - \sum^{N}_{j=1} \xi_j(\rho(k)) \left( K_{\zeta,j} (\zeta_j(k)-\zeta_j^*(k)) + K_{w,j}w(k) \right).
\end{equation}

The desired tracking values in the control law~\eqref{eq:ucomb} can be obtained using an approximated frozen model, where $\zeta_j^* = (x^*,0)$ and $ u_{\zeta}^* = u^*$, with the pair $(x^*(k),u^*(k))$ computed explicitly online, at each iteration given a desired conversion per pass $X_s$, as follows. Set $ x_1^*(k+1)  = x_1^*(k) = x_1(k)$ (since the dynamics are relatively slow) and  $x_2^*(k) = x_2(k)$ and  substitute $X_s$ and $x_1^*(k+1)$ into~\eqref{eq:SOCx} and\eqref{eq:SOCx1x2} to obtain $x_2^*(k+1) = ((1+\sqrt{x_1^*})/(1-Xs)-1)^2$ during charging or $x_2^*(k+1) = ((1-Xs)/(Xs+1/\sqrt{x_1^*}))^2$ when discharging. The corresponding ideal reference concentrations $(c_2,c_3,c_4,c_5)^*(k)$ can then be computed from $x_1^*(k)$, $x_2^*(k)$ using the procedure of Section~\ref{app:ideal_parameter}. Given $(c_2,c_3,c_4,c_5)^*(k)$, the reference varying parameters $(\rho_1,\rho_2,\rho_3,\rho_4,\rho_5)^*(k)$ can be computed via~\eqref{eq:rho},\eqref{eq:rho5}. Finally, the reference control input 
$u^*(k) = (x_2^*(k+1) - (1-\tau\rho_2^*(k)) x_2^*(k) - \tau \rho_4^*(k) I(k))/(\tau \rho_3^*(k))$ can be found via solution to~\eqref{eq:disc_ss}.

We note here, that this convex combination of vertices approach offers reduced computational complexity by sacrificing some performance (both in terms of conservativeness of the bounding convex polytope and accuracy of the reconstruction through linear combinations); however, this is desired as part of the design brief and will be demonstrated in Section~\ref{sec:implementation} as a justified investment, due to comparable performance. As an additional benefit, the analysis, in terms of system controllability and closed-loop stability, can be reduced to a problem parameterised by the vertices, which is naturally befitting, as shown in the following section.


\subsection{Closed-loop Analysis}\label{sec:analysis}
Performing the LPV embedding of Section~\ref{sec:lpv_model} rendered the system to be linear time-varying (since the parameter is time varying, i.e., $\rho = \rho(k)=\rho(x(k))$). Hence, to design a feedback controller that is stabilisable, as per Section~\ref{sec:param}, we are concerned with total controllability over the whole range of variation of $\rho$~\citep{WITCZAK2017729}. Thus, we need to test the controllability of the system~\eqref{eq:disc_ss} using the condition $\operatorname{rank} ( \mathcal{C}(k_0) )=\operatorname{dim}(x)$ for any $k_0$, where
\begin{equation}
\begin{split}
\mathcal{C}(k_0) &= \left[\begin{matrix} B(k_0) & A(k_0) B_{k_0+1} & A(k_0+1) A(k_0) B(k_0+2) & \cdots \end{matrix} \right.\\
			 &\left.\qquad\qquad \begin{matrix} \cdots & A(k_0+n-2) \cdot \ldots \cdot A(k_0+1) A(k_0)  B(k_0+n-1) \end{matrix} \right]  \\
			& =\begin{bmatrix}  	
\tau \rho_1(k_0) & \tau \rho_1(k_0+1)\\
\tau \rho_3(k_0) & (1 + \tau \rho_2(k_0)) \tau \rho_3(k_0+1)
\end{bmatrix}\!\!.
\end{split}
\end{equation}
Since the sampling period is $\tau>0$, and through consideration for the definition of $\rho$~\eqref{eq:rho} and its constraints (see~\eqref{eq:rho_constraint}), especially the away from zero range of operation of $\rho_1$, $\rho_2$ and $\rho_3$ (found experimentally), the matrix $\mathcal{C}(k_0)$ indicates total system controllability, i.e., for any $k_0$, $\operatorname{rank} ( \mathcal{C}(k_0) )=\operatorname{dim}(x)=2$.

With the controller feedback gains designed, as per Section~\ref{sec:param}, then, in order to implement either the online LQR controller~\eqref{eq:u}--\eqref{eq:Krhok}, or convex combination controller~\eqref{eq:Kcomb}--\eqref{eq:ucomb}, we require conditions to ensure the closed-loop system is stable. In the following, we employ a result from~\cite{McDeSe15} that provides conditions to assess robust system stability (in terms of boundedness of system trajectories) through the computation of invariant sets and apply it to the  control scheme of Section~\ref{sec:param}. In~\cite{McDeSe15}, an ``ultimate-bound''  set is derived that is asymptotically \emph{attractive}, i.e., the trajectories of the system ultimately converge to the set, and is \emph{invariant}, i.e., the trajectories cannot leave the set once inside. 

 Using~\eqref{eq:zeta} and \eqref{eq:u}, we have that $\zeta$ satisfies the dynamic equation
\begin{equation}
\label{eq:zeta+}
 \zeta(k+1) =\begin{bmatrix} A(\rho) - B(\rho) K_x(\rho) & -B(\rho) K_\sigma(\rho) \\ - \tau C(\rho) & 1 \end{bmatrix} \zeta 
+ \begin{bmatrix}  E(\rho) - B(\rho) K_w(\rho) & 0\\ 0 & \tau \end{bmatrix} \begin{bmatrix} w \\ r \end{bmatrix},
 \end{equation}
After a suitable reparameterisation of the indices in~\eqref{eq:zeta+} to fit the convex polytopic description (as in Section~\ref{sec:param}) and using the results of~\cite{McDeSe15} (details and discussion therein), provided a transformation $V$ exists\footnote{The existence of a transformation matrix, say $V_c$ with $\Lambda_c = \big|V_c^{-1} A_c V_c \big|$ being Schur (where $A_c$ denotes the closed-loop system), implies the existence of a common Lyapunov function (see~\cite{HaSe14}) and hence the stability of the closed-loop system.}
 such that
\begin{equation}
  \label{eq:Lambda}
  \Lambda := \max_{j,\ell \in \{1, \dots, N\}} \left |V^{-1} \begin{bmatrix} A_j - B_j K_{x_\ell} & -B_j K_{\sigma_\ell} \\ - \tau C_j & 1 \end{bmatrix} V \right|
\end{equation}
is a Schur matrix, then the trajectories of the closed-loop system~\eqref{eq:zeta+} are bounded and the set
\begin{equation}
  \label{eq:Szeta}
  \mathcal{S}_\zeta := \left \{ \zeta: |V^{-1} \zeta | \le (I-\Lambda)^{-1} \max_{j,\ell \in \{1, \dots, N\}} \left | V^{-1}  \begin{bmatrix}  E_j - B_j K_{w_\ell} & 0\\ 0 & \tau \end{bmatrix} \right | \begin{bmatrix} \overline{w} \\ \overline{r} \end{bmatrix} \right \},
  \end{equation}
is an attractive invariant set for the closed-loop system dynamics, where $|w| \leq \bar{w}$, $|r| \leq \bar{r}$. The system can then be suitably bounded as
\begin{equation}
\label{eq:zetabar} 
|\zeta| \leq \overline{\zeta} :=  |V|(I-\Lambda)^{-1} \max_{j,\ell \in \{1, \dots, N\}} \left | V^{-1}  \begin{bmatrix}  E_j - B_j K_{w_\ell} & 0\\ 0 & \tau \end{bmatrix} \right | \begin{bmatrix} \overline{w} \\ \overline{r} \end{bmatrix}.
\end{equation}
The existence of the transformation $V$ such that
the matrix $\Lambda$~\eqref{eq:Lambda} is Schur, ensures closed-loop system stability (in terms of boundedness) of the overall system using the tracking controller with integral action~\eqref{eq:u} or~\eqref{eq:ucomb}. Note that tighter invariant sets for the system dynamics, where necessary (e.g. when used in fault detection mechanisms), can be obtained by considering a ``central system'' (see, e.g.,~\cite{McDeSeIJC17}, which considers the constant offset of the reference, $r$, as the centre of its containing set). Furthermore, the above stability analysis can also be extended to consider the tracking error dynamics or when additional bounds are known with respect to measurement and process noise.

\section{Implementation Results}
\label{sec:implementation}
In this section, the simulation and experimental results of the proposed LPV based control scheme (see Section~\ref{sec:controller}), are presented. Details of the system parameters are given in Table~\ref{tab:sys_parameters}.

\begin{table}
\centering
\begin{tabular}[t]{lccc}
\hline
&Value& Units&Description\\
\hline
$L_{pe}$		&3				        &dm                     &porous electrode length\\
$W_{pe}$		&0.03				    &dm                     &porous electrode width\\
$H_{pe}$		&2				        &dm                     &porous electrode height\\
$n$			    &1				        & -/-                   &number of electrons transferred\\
$F$             &96485                  &C mol$^{-1}$           &Faraday's constant\\
$\bar{c}$		&1.6				    &mol L$^{-1}$           &total vanadium concentration\\
$c_{max}$       &1.44                   &mol L$^{-1}$           &maximum half cell concentration\\
$c_{min}$       &0.16                   &mol L$^{-1}$           &minimum half cell concentration\\
$M$			    &9				        &cells                  &number of cells in the stack\\
$E'_0$		    &1.4				    &V                      &formal potential\\
$R$		        &8.314				    &J mol$^{-1}$ K$^{-1}$  &gas constant\\
$\frac{k_2}{d}$	&$3.17\times10^{-7}$	&dm s$^{-1}$         &V$^{2+}$ diffusivity coefficient / membrane thickness\\
$\frac{k_3}{d}$	&$7.16\times10^{-8}$	&dm s$^{-1}$         &V$^{3+}$ diffusivity coefficient / membrane thickness\\
$\frac{k_4}{d}$	&$2\times10^{-7}$		&dm s$^{-1}$         &V$^{4+}$ diffusivity coefficient / membrane thickness\\
$\frac{k_5}{d}$	&$1.25\times10^{-7}$	&dm s$^{-1}$         &V$^{5+}$ diffusivity coefficient / membrane thickness\\
$V_t$			&3.88				    &L                      &half cell electrolyte volume\\
$I_{max}$		&30				        &A                      &maximum charging current\\
$I_{min}$		&-30				    &A                      &maximum discharging current\\
$Q_{max}$		&0.0286			        &Ls$^{-1}$              &maximum electrolyte flow rate\\
$Q_{min}$		&0.013			        &Ls$^{-1}$              &minimum electrolyte flow rate\\
$T$			    &293.15			        &K                      &ambient temperature\\
\hline
\end{tabular}
\caption{Pilot VRB System Parameters}
\label{tab:sys_parameters}
\end{table}

\subsection{Simulation Results}
\label{sec:sim_results}
 The initial condition of the VRB, is determined by an initial state of charge, $SOC(0)$, and corresponding concentrations, given by
\begin{equation}
\label{eq:c0}
c_{2,5}(0) = \bar{c}SOC(0), \qquad c_{3,4}(0) = \bar{c}(1-SOC(0)),
\end{equation}
where $SOC(0) = 0.1$ for the charging and $SOC(0) = 0.9$  for discharging (i.e., $10\%$ and $90\%$ initial respective state of charge)
Note, equal concentrations for the tank and the cell implies a zeroed initial conversion factor, $X(0) = 0$, since the state of charge at the inlet is equal to the state of charge at the outlet. The initial state, $x(0) = (x_1(0),x_2(0))$, is then given by 
\begin{equation}
\label{eq:x0}
x_1(0) = x_2(0) = \frac{c_2(0)c_5(0)}{c_3(0)c_4(0)}.
\end{equation}

The current input, $I$, is considered as a non-ideal charging current with $25\%$ fluctuation and is implemented as a pseudo-random square wave
\begin{equation}
\label{eq:I}
I = (1+k_i) I_s,
\end{equation}
with nominal current, $I_s = 20A$ for charging and $I_s = -20A$ for discharging, with random variation, $k_i \in \begin{bmatrix} -0.5 & 0.5 \end{bmatrix}$, sampled every $600s$. 
All other simulation parameters and constraints are as in Table~\ref{tab:sys_parameters}.

Stabilisability (via controllability tests, see, e.g., \cite{McCloyDeDonaSeron18}) was confirmed for each pair of system vertices $(A_{\zeta,j},B_{\zeta,j})$ for $j =1\dots,N$. The controller gains are then obtained via~\eqref{eq:cost}, with $Q(\rho) = Q_s = \operatorname{diag}\{1,1,5\times10^3\}$, $R(\rho) = R_s = 1\times10^4$ designed to place significant importance on the tracking performance, $X\to X_s$, i.e., operational efficiency, whilst not neglecting the pump energy losses due to high flow rates. The existence of the transformation $V$ such that the matrix $\Lambda$ in~\eqref{eq:Lambda} is Schur was verified, which ensures closed-loop system boundedness of the overall system using the tracking controller with integral action~\eqref{eq:u}--\eqref{eq:ucomb}.

\begin{figure}
\centering
  \includegraphics[width=\linewidth]{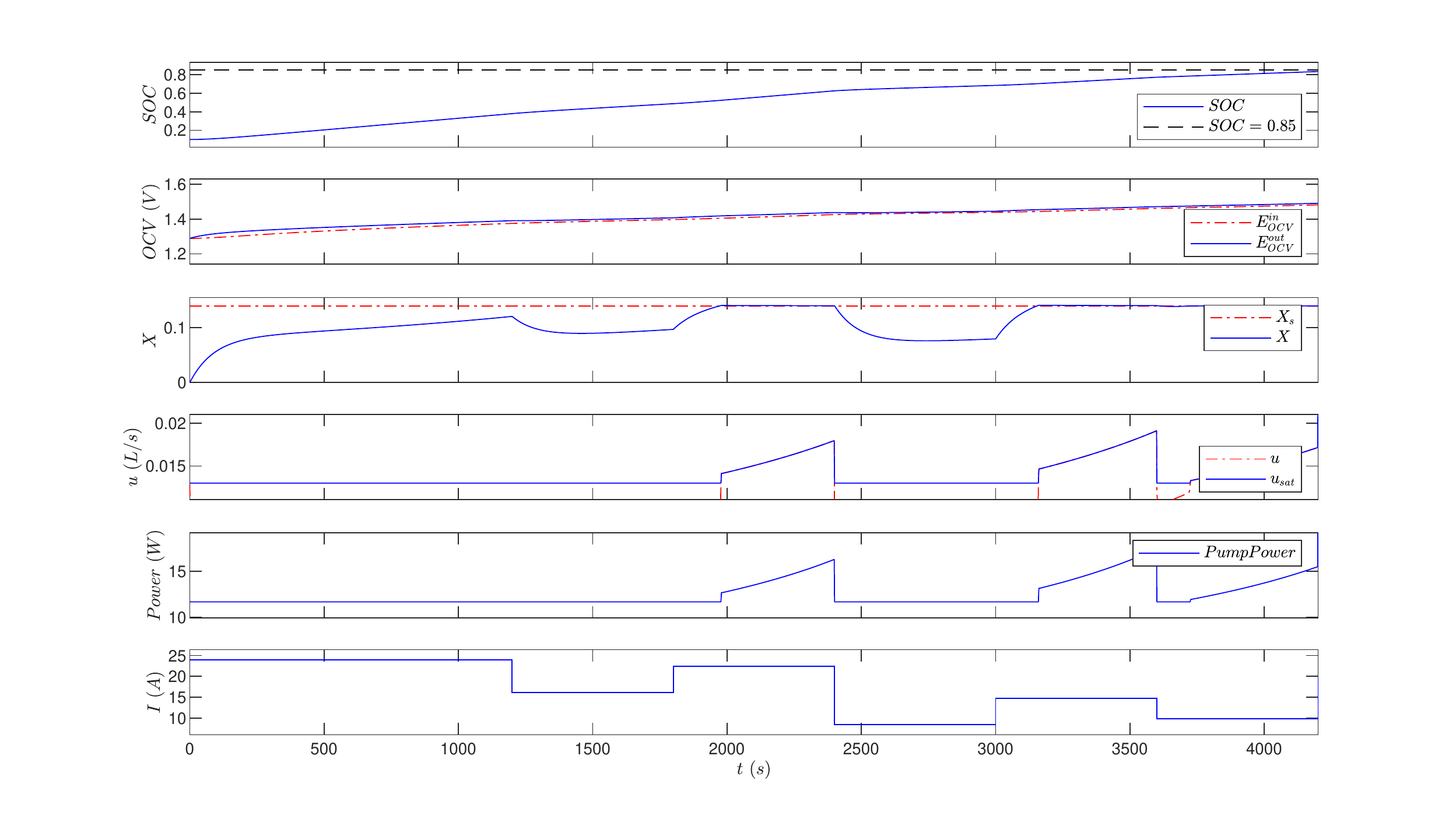}
  \caption{VRB Charging Simulation using Convex Combination Feedback Gains}
  \label{fig:simPlotLPVKcv}
\end{figure}

\begin{figure}
\centering
  \includegraphics[width=\linewidth]{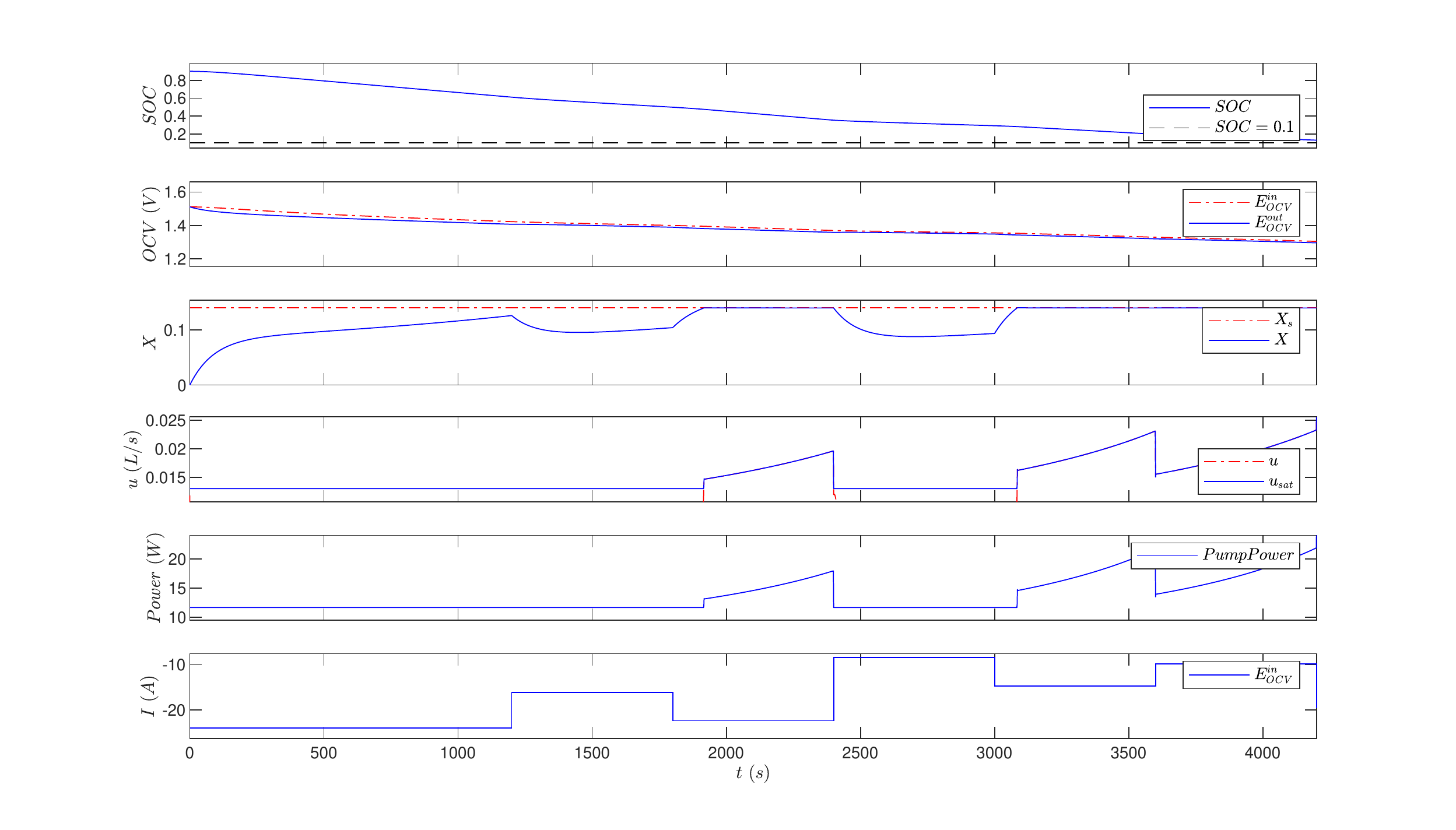}
  \caption{VRB Discharging Simulation using Convex Combination Feedback Gains}
  \label{fig:simPlotLPVKdv}
\end{figure}

We simulated the proposed control scheme (as shown in Figure~\ref{fig:block_diagram}) to achieve a conversion factor setpoint of $X_s = 0.14$, whilst charging the VRB from $10\%$ to $85\%$ SOC. The controller was simulated with the feedback gains computed using a convex combination~\eqref{eq:Kcomb},~\eqref{eq:Kvert} as shown in Figure~\ref{fig:simPlotLPVKcv}. 
The VRB was then simulated for the discharging scenario of $85\%$ to $10\%$ SOC, as can be seen in Figure
~\ref{fig:simPlotLPVKdv}. As shown, the proposed control scheme is capable of charging and discharging the VRB under a fixed conversion factor when physically viable (i.e., considering SOC under applied current supply/load), subject to fluctuations in charging and discharging current. Note that during the charging simulation, a decrease in charging current corresponds to a decrease in conversion factor for the same SOC (and similarly for discharging), which additionally requires a reduced flow rate to maintain the desired conversion per pass (a known characteristic of VRB operation). Likewise, an increase in charging current for the same SOC will result in a higher conversion per pass and require an increased flow rate to maintain the desired conversion factor setpoint (and similarly for discharging). When the pump flow rate was not saturated, the desired conversion factor was suitably maintained as demonstrated in Figures \ref{fig:simPlotLPVKcv} and \ref{fig:simPlotLPVKdv}. As also shown, the conversion per pass was able to vary freely when the electrolyte pump is not capable of delivering the required flow rate determined by the controller due to: (1) lower saturation -- the minimal electrolyte flow rate to maintain effective battery operation; (2) upper saturation -- the physical maximal flow rate deliverable by the pump. In addition, for the second scenario, higher SOCs alongside relatively high charging current (and similarly for low SOC with high discharging current), for which the flow rate is insufficient, would require the introduction of a current limiting module to protect the battery, as discussed in Section \ref{sec:discussion}.   

\begin{figure}
\centering
  \includegraphics[width=\linewidth]{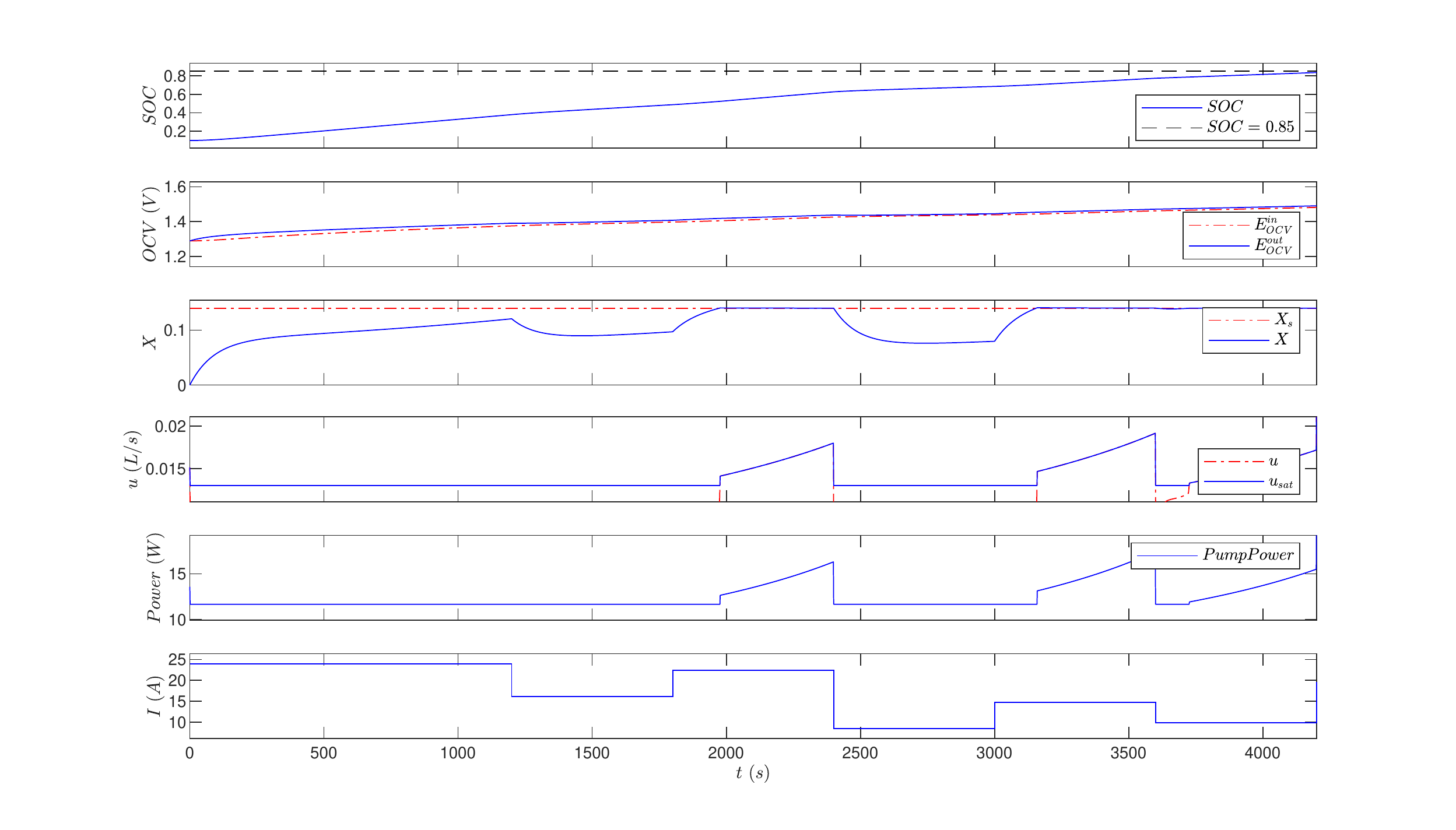}
  \caption{VRB Charging Simulation using Online LQR Feedback Gains}
  \label{fig:simPlotOLPVKcv}
\end{figure}

 For comparison, we also present the alternative (more computationally complex) controller implementation, whereby the controller feedback gains are computed by solving the algebraic Riccati equation at each time step~\eqref{eq:Krhok}, with results shown in Figure~\ref{fig:simPlotOLPVKcv}.
Comparing the online LQR~\eqref{eq:Krhok} results of Figure \ref{fig:simPlotOLPVKcv} with the proposed convex combination~\eqref{eq:Kcomb} results of Figure \ref{fig:simPlotLPVKcv}, we observe that there is no significant improvement, despite the increased computational burden. The tracking performance of the proposed method is almost identical, with a very minimal  increase in control magnitude (and hence pump power consumption) in response to aggressive changes in charging current (see $t\approx 1980 s$, $t \approx 3160 s$) relative to the online LQR controller method. We also conducted simulation studies with fixed feedback gain control approaches using a fixed linearised model (about a target conversion per pass and SOC), including both traditional state feedback and PI-based designs and found those approaches completely ineffective in controlling the electrolyte conversion per pass, through an effective range of SOC and under variable supply or load (not illustrated here), due to the nonlinearity of the battery dynamics. For this reasoning, the experimental implementation is completed using the parameter varying feedback gains computed via a convex combination of vertices.

\subsection{Experimental Results}
\begin{figure}
\centering
  \includegraphics[width=0.8\linewidth]{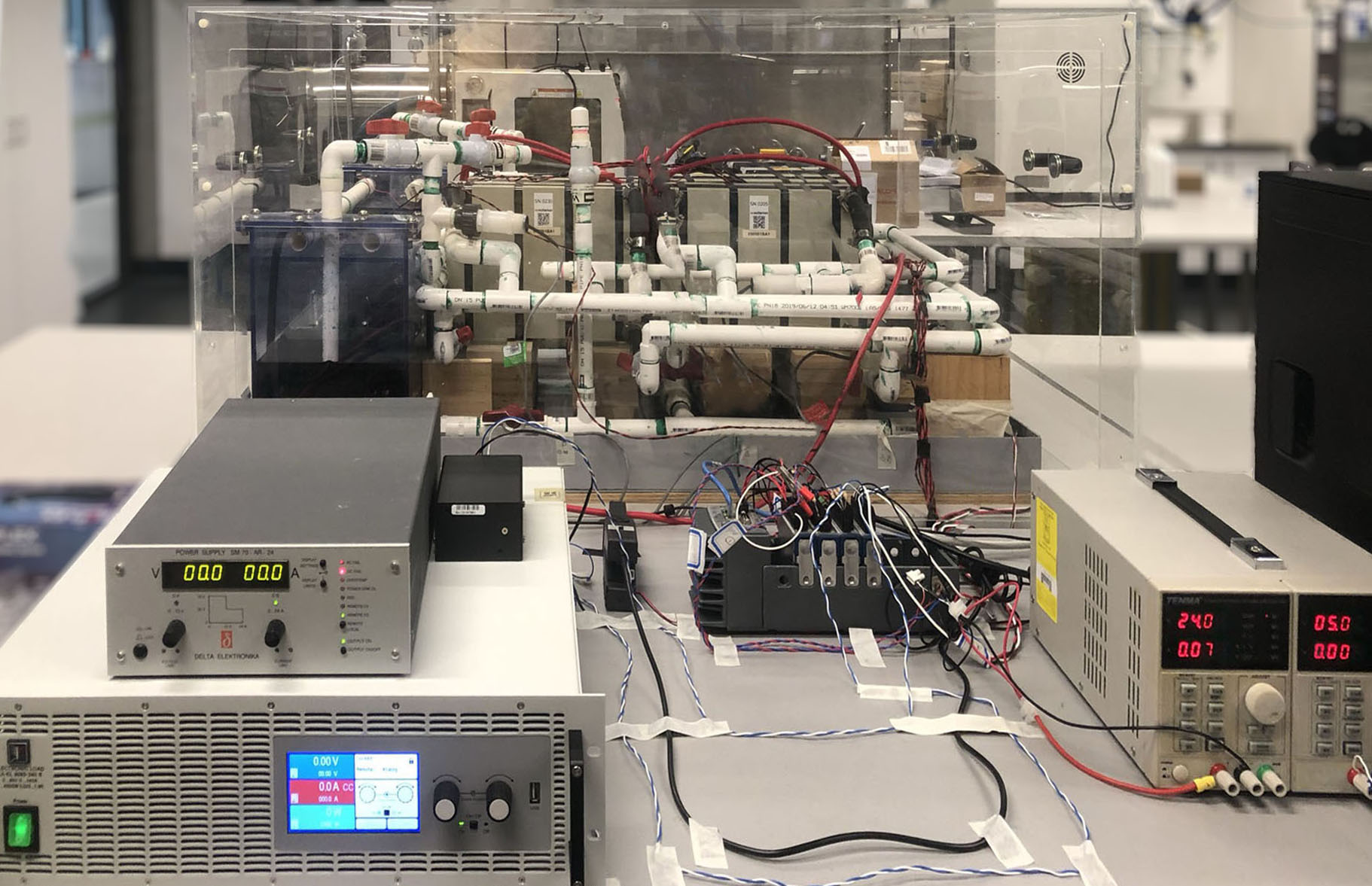}
  \caption{Pilot VRB Experiment}
  \label{fig:pilot_VRB}
\end{figure}

In this section, the LPV controller developed in Section~\ref{sec:controller} and simulated in Section~\ref{sec:sim_results} is implemented on the laboratory scale VRB system shown in Figure~\ref{fig:pilot_VRB}, with parameters as in Table~\ref{tab:sys_parameters}. In summary, the main control user interface is written in LabVIEW; the controller is compiled and stored on a CompactRIO data-acquisition module, which directly interfaces with the physical battery, by sending voltage signals, $V_p$, to the electrolyte pumps, and taking OCV measurements, which are relayed back to LabVIEW. The relationship between the electrolyte flow rate, $Q$, and the required pump voltage, $V_p$ is described using the following linear expression
\begin{equation}
V_p = m_p Q + b_p, \qquad V_{p,min} \leq V_p \leq V_{p,max}
\end{equation}
where, via experimental calibration, $m_p = 1.838$, $b_p = 1.743$ and $V_{p,min} = 3.176$, $V_{p,max} = 4.892$.
In addition, a ``shut-off'' or ``maintenance'' controller is switched in at $\geq 90\%$ (high) or $\leq 10\%$ (low) state of charge to avoid the adverse affects of very high and low state of charge in the VRB. A pseudo-random charging current was applied to imitate a variable supply from renewable sources.

\begin{figure}
\centering
  \includegraphics[width=\linewidth]{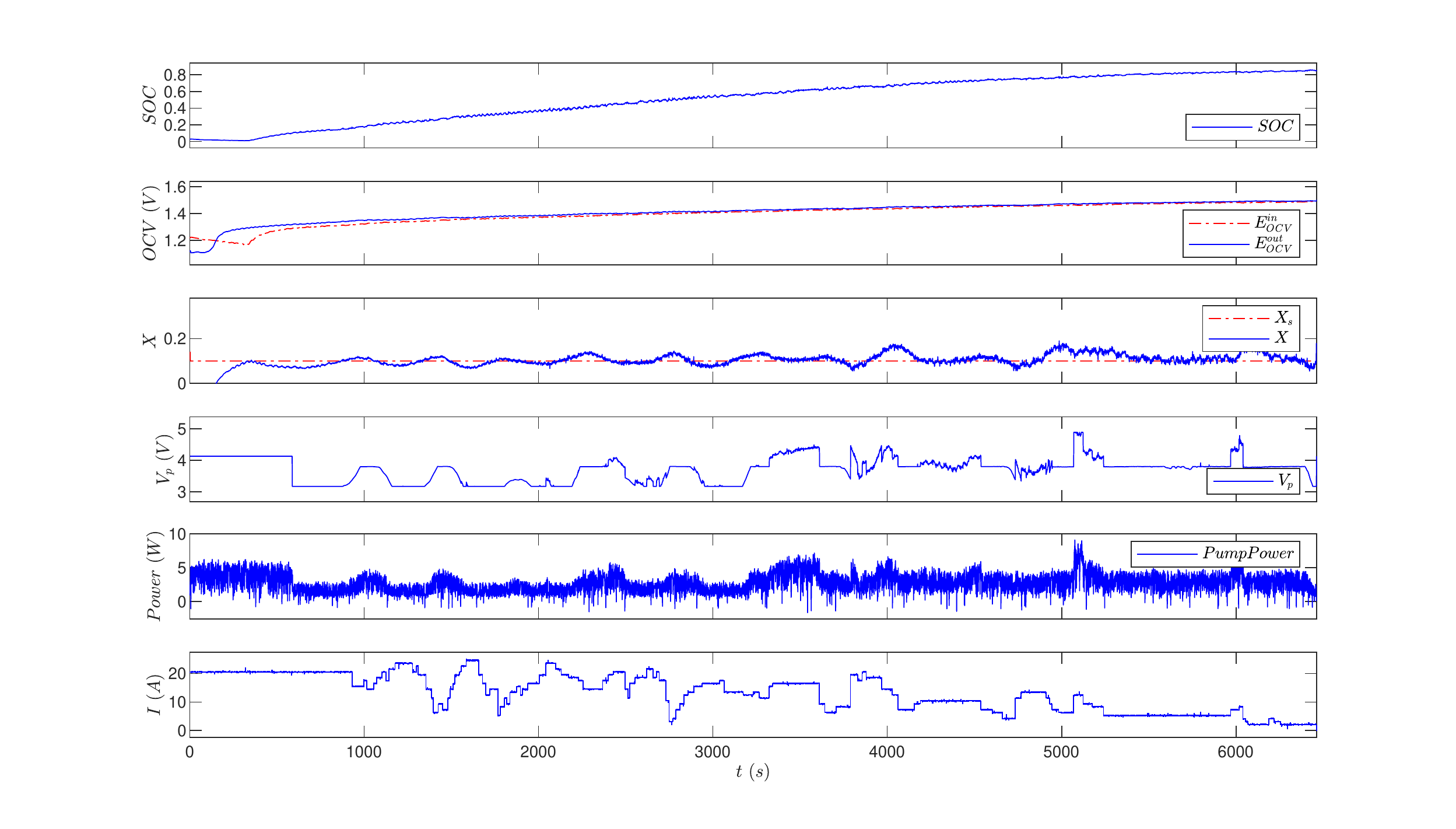}
  \caption{Pilot VRB Charging Experiment using Convex Combination Feedback Gains}
  \label{fig:expPlotLPVKcv}
\end{figure}

We implemented the overall scheme on the pilot VRB laboratory setup to achieve a target conversion factor of $X_s = 0.1$, whilst charging the VRB from $1\%$ to $85\%$ state of charge, see Figure~\ref{fig:expPlotLPVKcv}. As shown, the overall control scheme is capable of charging the VRB under a fixed conversion factor (when physically viable), subject to aggressive fluctuations in charging current. 
Further discussion of the controller implementation is provided in the following Section.

\subsection{Discussion}
\label{sec:discussion}
Within the current framework, the parameter varying feedback gains could be computed both online by solving the algebraic Riccati equation at each time step \eqref{eq:u}, or, by implementing the feedback gains as a convex combination \eqref{eq:ucomb}, i.e., at each sampling instant, computing the gains as in~\eqref{eq:Krhok}, or as in~\eqref{eq:Kcomb},~\eqref{eq:Kvert}. An early design choice was made, due to hardware limitations, to significantly reduce the online computational burden with minimal performance loss by implementing the feedback gains as proposed in Section \ref{sec:controller} (see, specifically,~\eqref{eq:ucomb}). For this same reasoning, alternative approaches (such as online extended LQR~\cite{SinghPal17} and MPC~\cite{RaMa09}) were also found unsuitable due to their high computational burdens. It is also important to note that computationally simpler approaches, such as traditional PI-based designs, were completely ineffective in achieving the desired conversion factor, in both simulation and experiment, in the presence of variable supply or load. This can largely be attributed to a combination of the nonlinear nature of the process and sensor accuracy. 

As discussed in Section~\ref{app:ideal_parameter}, the proposed control scheme currently relies on the ideal model for extracting varying parameter measurements, i.e., that the two half cells are fully balanced from tank to cell (completed manually during calibration). If this relationship doesn't hold precisely, it can result in performance inefficiencies due to incorrect computation of the design parameters (dependent on $\rho$ and hence the concentrations). The proposed framework additionally offers some robustness to such modelling uncertainties (see also, e.g., the approach in~\cite{NazariSeronDeDona15} for handling varying parameter uncertainty), as demonstrated by the laboratory experiment. Future work may consider additional monitoring and correction options for half-cell and tank/cell balance and potential alternative measurement strategies, including nonlinear estimation based approaches and improved sensor configurations.

The proposed LPV based control framework offers significant performance improvement over existing (fixed gain) methods whilst offering minimal computational burden. A significant advantage of the proposed approach is that the (non-unique) LPV embedding facilitates a stabilisable system description in the widely available and well-studied state space structure, which to the authors' knowledge at the time of publication was yet to be realised for the VRB electrolyte flow model. This contribution consequently permits application of a wide range of control design tools via the state space system description. 

In this study, the charging/discharging current is treated as a disturbance, but needs to be constrained in practice. For example, charging current should be under a limiting current which is a function of the flow rate and SOC to avoid gassing side reaction \cite{Akter2019}. Further work includes treating the charging/discharging current as an additional manipulated variable to  extend the proposed approach to optimise the economic benefit of battery system operations. 

\section{Conclusion} \label{sec:conclusion}
In this article, the linear parameter varying framework facilitated system modelling and control of a VRB system to achieve a desired conversion per pass, leading to  efficient battery operation. The dynamic equations for the VRB system were first embedded in an LPV description via new state definitions and reparameterisation of system nonlinearities. The system model was then discretised and augmented to include an integral state, such that steady state errors with respect to the desired conversion fraction could be eliminated.  For each of the charging and discharging scenarios, a performance output model for the conversion per pass as a function of the states was proposed and combined with a set of convex polytopic state feedback tracking controllers. The overall control scheme was simulated subject to fluctuations in current, and illustrated the available design trade-off between conversion rate tracking performance and power consumption of the pump. Due to the relatively low complexity of computing the convex combination of vertices online, for the control input, the proposed approach offers ease of implementation and low hardware requirements for battery control and management applications. The proposed control scheme was successfully implemented on a laboratory setup. 

\section*{Acknowledgements}
This work was partially supported by Australian Research Council Industrial Transformation Research Hub for Integrated Energy Storage Solutions IH180100020. The authors would also like to acknowledge Mr Longgang Sun's assistance in preparing and conducting the VRB experiments.

\bibliography{VRB_LPV_overleaf}

\end{document}